\documentclass[submitting]{nst}

\usepackage{subfigure,dcolumn}
\usepackage{epstopdf}
\usepackage{mhchem}
\usepackage{upgreek}
\usepackage{hyperref}

\begin{document}

\title{Proposal for Measurement of the Two-body Neutron Decay using  Microcalorimeter\mbox{}}
\thanks{Project supported by the National major scientific research instrument development project (Grant No.11927805)}

\author{Shuo Zhang}
\email[Corresponding author, ]{shuozhang@shanghaitech.edu.cn}
\affiliation{Center for Transformative Science, ShanghaiTech University, ShangHai, 201210, China}

\author{Xavier Mougeot}
\affiliation{CEA, LIST, Laboratoire National Henri Becquerel, CEA-Saclay, Gif-sur-Yvette Cedex,91191, France}

\author{Song-Lin Wang}
\affiliation{Institute of High Energy Physics, Chinese Academy of Sciences, BeiJing, \quad 100049, China}

\author{Jian-Rong Zhou}
\affiliation{Institute of High Energy Physics, Chinese Academy of Sciences, BeiJing, \quad 100049, China}

\author{Wen-Tao Wu}
\affiliation{Shanghai Institute of Microsystem and Information Technology, Chinese Academy of Sciences, ShangHai, 200050, China}

\author{Jing-Kai Xia}
\affiliation{Center for Transformative Science, ShanghaiTech University, ShangHai, 201210, China}

\author{Rui-Tian Zhang}
\affiliation{Institute of Modern Physics, Chinese Academy of Sciences, LanZhou, \quad 730000, China}

\author{Le Zhang}
\affiliation{School of Physics and Astronomy, Sun Yat-Sen University, Guangzhou, 510297, China}

\begin{abstract}
The bound beta-decay (BoB) of neutron is also known as the two-body neutron decay, which is a rare decay mode into a hydrogen atom and an anti-neutrino. The state of neutrino can be exactly inferred by measuring the state of hydrogen atom,  providing a possible pathway to explore new physics. However, this rare decay mode has not yet been observed so far since it was predicted in 1947. The challenge in observing this decay is not only that its cross section is extremely low, equivalent to about branching ratio of the order of $10^{-6}$ of the three-body decay, but also that the final-state hydrogen atom is neutral and has extremely low kinetic energy, which cannot be effectively detected. In this study, we propose a microcalorimeter-based scheme for measuring the kinetic energies of hydrogen atoms produced from BoB of ultracold neutrons, which has a great advantage in terms of accuracy of the energy measurement. In this study, first, several important issues that require rigorous considerations for the decay measurements and possible solutions are discussed. Then, the requirements of the neutron flux and the appropriate structure design of the microcalorimeter are present by theoretical calculations. In short, this paper outlines our proposed novel experimental scheme for observing the BoB mode, addressing the possible solutions to all the necessary problems.


\end{abstract}


\maketitle

\section{Introduction}\label{sec.I}

\begin{figure}[!htb]
\includegraphics[width=.8\hsize]{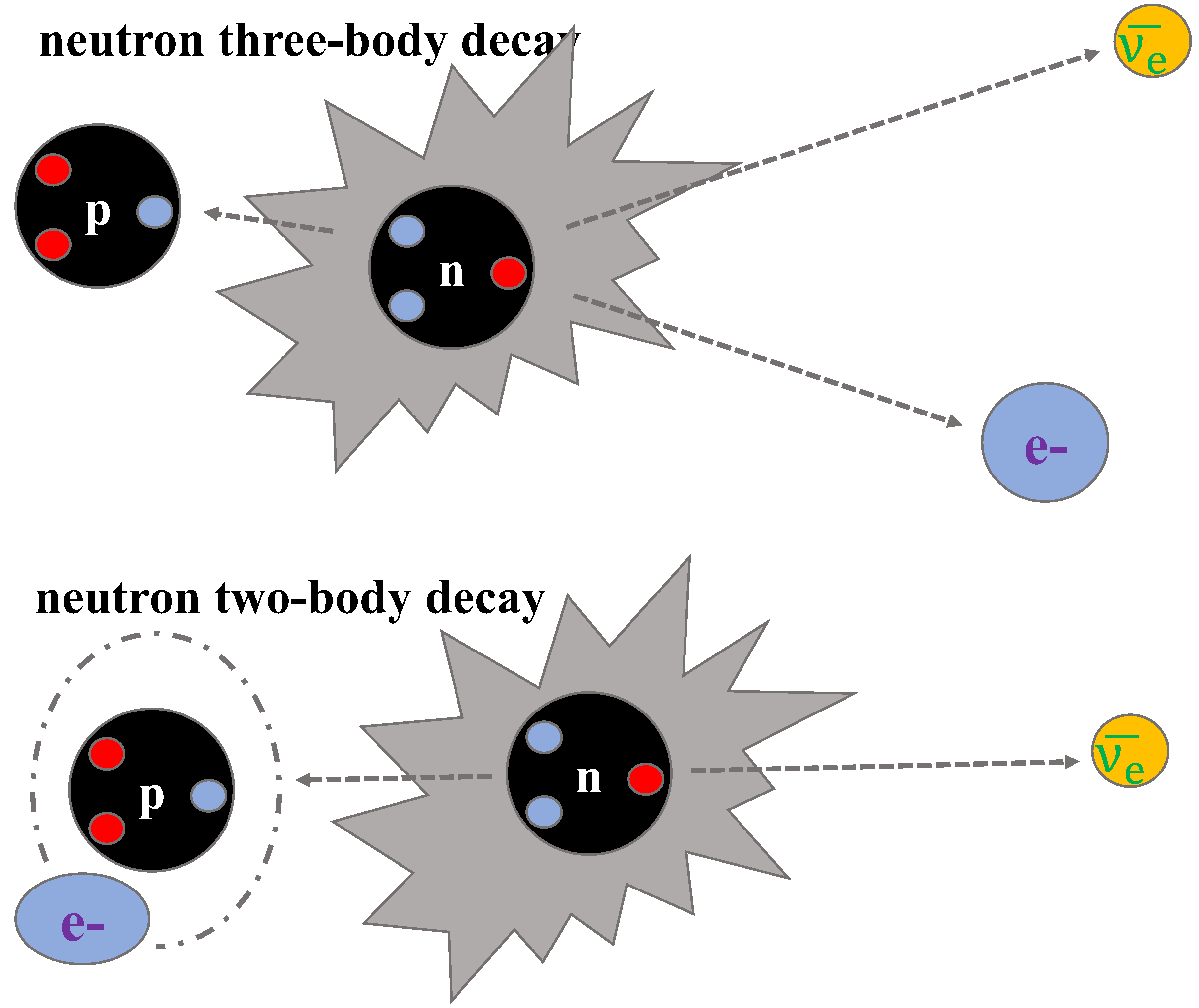}
\caption{Comparison of the two decay modes: 1)  a regular three-body decay mode of the neutron into protons, electrons and neutrinos, with the continuum energy spectra for all three; 2) an exotic decay mode of the neutron into a hydrogen atom and an anti-neutrino, which has a extremely small cross section and the final state particles are both monoenergetic.}
\label{fig1}
\end{figure}

Neutron decay is closely related to fundamental physics. The fact that, no new particles other than the Higgs boson have been found at the LHC, has prompted a turn to the search for signatures of new physics in the low and medium-energy regions~\cite{1-BoB-th4}. Experiments related to $\beta$ decay are an important avenue to explore the standard model of electroweak interactions, e.g.,  a neutron decaying into a proton, an electron, and an electron antineutrino, which is a concise model for studying weak interactions. In 1947 Daudel, Jean and Lecoin~\cite{1-BoB-th5,1-BoB-th6} have predicted the existence of a two-body $\beta$-decay mode where the daughter nucleus and the electron remain bound. In the case of the free neutron, the neutron two-body decay mode is $\rm{n \rightarrow H+\bar {\nu}_e}$, which is also referred to as "bound $\beta$-decay" (BoB). As this decay results in a two-body final state, it is theoretically quite straightforward to determine the states of individual particles. Many theoretical studies have been carried out on this two-body decay~\cite{1-BoB-th6,1-BoB-th1}, although it has never been observed in experiments so far. The successful measurement of this process is of great physical importance~\cite{1-BoB-th1}. As the BoB leads to a two-body final state, the spin state of the anti-neutrino is thus mirrored by the outgoing hydrogen atom. As such, measuring the hyperfine spin state of the hydrogen atom would therefore contain full information relating to the momentum direction of the antineutrino~\cite{McAndrew:2014iia}. Moreover, the measurement of the BoB branching ratio can also be a solution to the recently proposed neutron lifetime puzzle~\cite{1-BoB-th8,Czarnecki:2018okw}.

 The major challenges in observing the BoB decay and investigating its properties lie in the small predicted branching ratio $\sim4\times10^{-6}$  of the dominated three-body decay mode~\cite{1-BoB-th9,1-BoB-th8,1-BoB-th10}, together with the detection of low-energy electrically neutral hydrogen atoms in the final state. According to the theory, the hydrogen atom in the final state has a kinetic energy of 325.7 eV, corresponding to a velocity $\sim 10^5 {\rm m/s}$, and is expected to be populated with zero angular momentum, specifically, with $83.2\%$ of atoms in the 1s-state and $10.4\%$ in the 2s-state and the remainder in an $n$s-state where $n\geq3$~\cite{1-BoB-th2}.

In this study, we propose a scheme to detect the BoB that will rely on the  measurement of H atoms through its energy spectrum structure, in order to observe for the first time this hitherto unobserved decay mode. As known, the standard three-body decay mode is $n \rightarrow H^{+}+e^{-}+\bar {\nu}_{e}$, where the proton and electron are charged particles that are easily captured by solid materials, while the antineutrino is extremely difficult to capture. The total energy in the three-body decay is conserved, leading to continuum energy spectra for each of particles. Since the mass of the proton is much larger than that of the electron and neutrino, the decay energy is obtained mainly by the electron and neutrino, both of which have kinetic energies in the range of 0--782 keV, while the energy spectrum of the proton lies in the range of 0--750 eV~\cite{1-BoB-th3}.

\begin{figure*}[!htb]
\includegraphics[width=0.9\hsize]{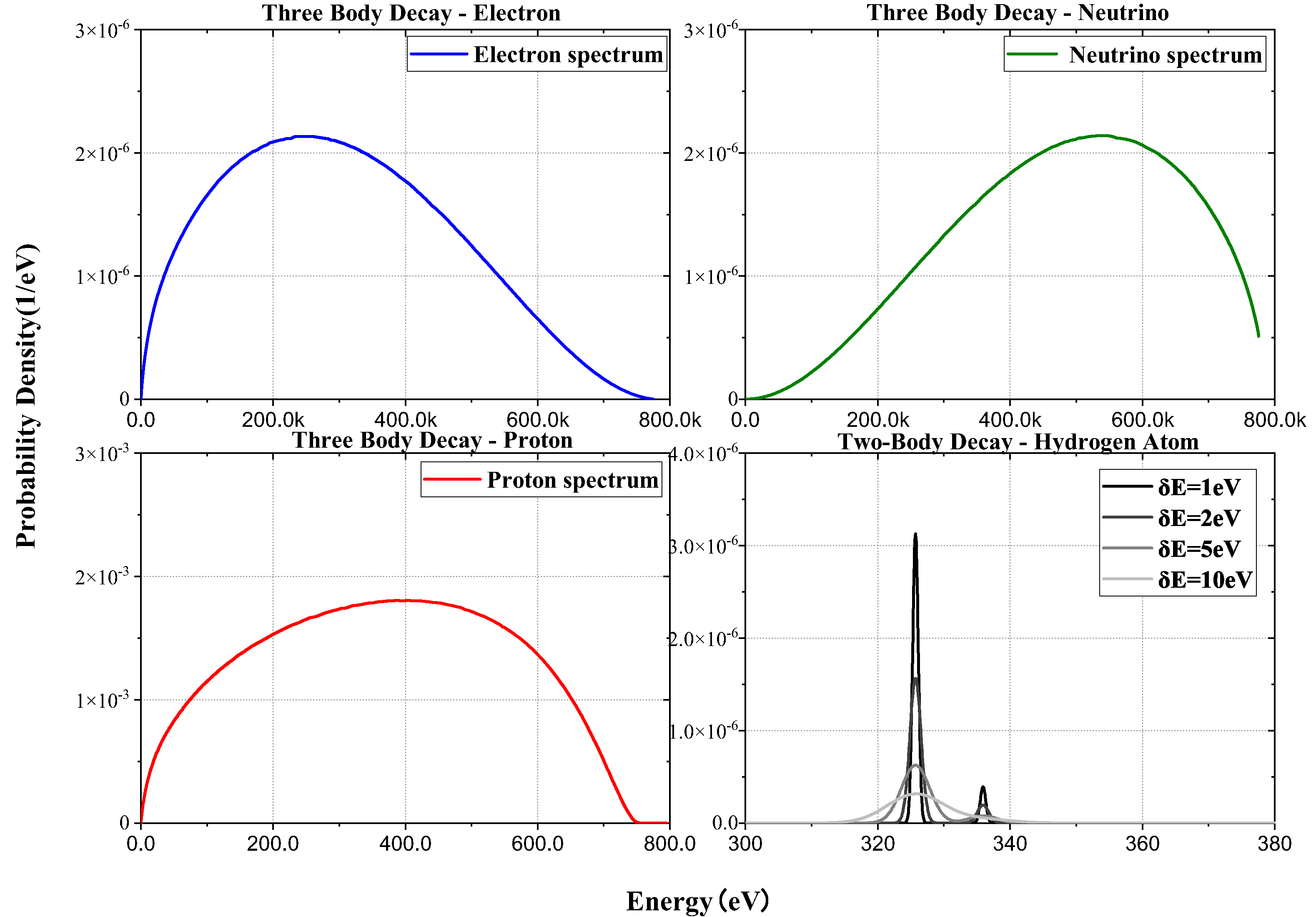}
\caption{Comparison of the probability densities  of electrons (upper left), neutrinos (upper right) and protons (lower left) produced by neutron three-body decay and hydrogen atoms (lower right) by two-body decay~\cite{1-BoB-th3}. The broadening of the hydrogen atom energy spectrum is determined by the initial kinetic energy of the parent neutron. With varying the initial kinetic energy, the hydrogen atom energy spectra broadened to 1 eV, 2 eV, 5 eV, and 10 eV are shown. Note that, the predicted probablity density of hydrogen atoms is highly suppressed by the two-body-decay branching ratio that is assumed to be $4\times 10^{-6}$.}
\label{fig2}
\end{figure*}

\begin{figure*}[!htb]
\includegraphics[width=0.9\hsize]{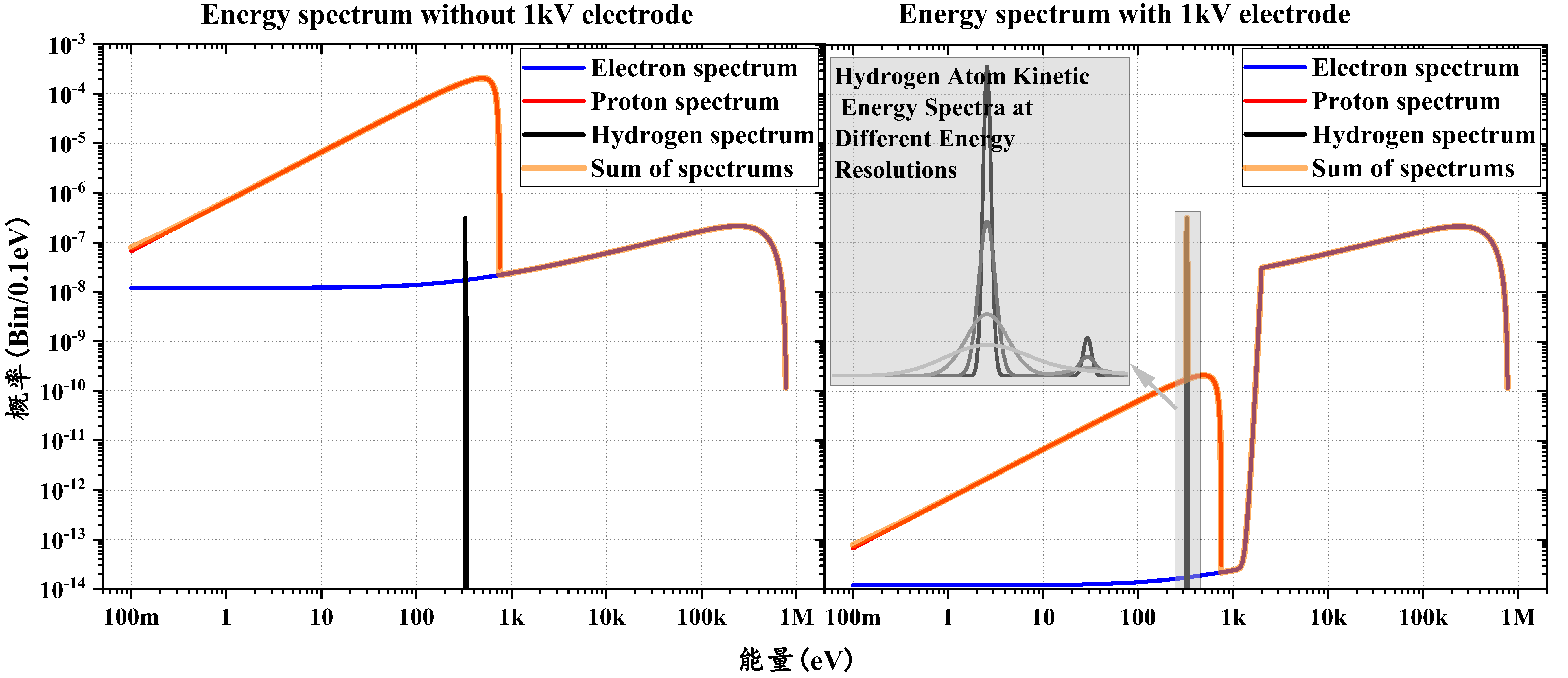}
\caption{Same as Fig.~\ref{fig2}, but for comparison of the BOB-induced spectral structures before (left) and after (right) electric field screening. Adding a voltage of 1 kV or more can significantly eliminate the effects of electrons and protons. The protons can be completely eliminated and are thus invisible on the plot, since their kinetic energies are less than 1 keV.}
\label{fig3}
\end{figure*}

For the BoB mode, the conservation of momentum and energy during the decay leads to the energy carried by the antineutrino is constant, while the kinetic energy of the hydrogen atom is also constant (325.7 eV),
making it easy to be captured by solid materials. Since the hydrogen atom in the $n$s ($\geq 2$) state after the two-body decay can be captured by the detector, it will de-excite to the 1s state very quickly, transferring its thermal energy greater than 10.2 eV to the detector. Thus, the kinetic energy spectra of the hydrogen atoms on the detector would be a multiple line structure dominated by 325.7 eV and 335.9 eV. The lower right panel of Fig.~\ref{fig2} shows only these two spectral lines, and the other spectral lines from $n$s states are ignored for the moment. Moreover, three-body-decay protons will introduce a strong background continuum spectrum contaminating the measurement of the hydrogen atoms, which needs to be removed to obtain a high signal-to-noise ratio. Since the charged protons and electrons are susceptible to an external electric field, one can set up positive and negative electrodes at about 1 kV in a low-temperature vacuum, resulting in a large difference in signal amplitudes between electrons with  $E>1$ keV and hydrogen atoms at 325.7 eV, which therefore can be easily identified. In Fig.~\ref{fig3}, the expected shapes of energy spectrum before and after electric field screening is shown. As seen, combine with the 1kV screening electric field, the central energy of the electrons is in the energy region of several hundred keV,  which is very different from the line signature of the two-body-decay hydrogen atoms.

A microcalorimeter is a sensitive detector for heat signals, with extremely high energy resolution at eV level and no dead layer, as well as a wide range of available absorption materials~\cite{2-uCal-theory,3-basic-theory}. The energy of the particle deposition is converted into heat and the resulting temperature rise is measured. Since hydrogen atoms, electrons and protons can all generate heat signals at the detector, and the energy resolution at the eV level fully satisfies the requirement for resolution of a single energy peak at 325.7 eV, the microcalorimeter detector is thus an ideal detector for measuring the BoB according to energy spectrometry. Although electrons and protons below 1 keV can be completely eliminated by the electric field, there are still some background processes that would produce a continuum at the detector. As seen from the lower right panel of Fig.~\ref{fig2}, if the initial kinetic energy of neutrons is much less than 1 eV and the resolution of the detector is around 1 eV, not only can the hydrogen atom be clearly seen in the 1s or 2s or even a higher state, but also a high signal-to-background can be obtained. For the transition edge sensor (TES)-based microcalorimeter, in the 1 keV energy range, the best reported energy resolution is about 0.75 eV~\cite{4-TES0.75eV}, and the best one we obtained so far is about 1.4 eV~\cite{5-TES1.4eV}, so that the detector performance is basically appropriate for the BoB measurement. More importantly, as the energy range of interest is around 325.7 eV, a better energy resolution can be obtained by further reducing the heat capacity of the microcalorimeter~\cite{3-basic-theory}.

This paper is organized as follows. In Sect.~\ref{sec.II}, we will present the theoretical calculations for the BoB and summarize various neutron sources, then discuss the the requirements for the BoB measurements, in addition we briefly review the working principle of microcalorimeter. In Sect.~\ref{sec.III}, we outline our experimental concept and investigate suitable neutron sources, as well as discuss measurement related issues, such as the $^3$He problem in refrigerators, the elimination of proton effects, effects of high-energy particles, the method of readout electronics, etc. Finally, we draw our summary and conclusions in Sect.~\ref{sec.IV}.

\section{basic principles}\label{sec.II}

\subsection{Theoretical Calculations of Neutron Two-Body Decay}\label{A}

Let us consider the BoB process, ${\rm n} \rightarrow {\rm H}+\bar{\nu}_{\rm e}$, where the expression for the kinetic energy of the hydrogen atom can be derived from the conservation of energy and momentum. In the following, the subscripts of n, H, $\bar{\nu}_e$ represent neutron, hydrogen atom and electron antineutrino, respectively. $E$, $P$ and $M$ denote the energy, the momentum and the rest mass, respectively. We will use natural units in which $c = 1$ throughout. First, the conservation of energy leads to
\begin{equation}\label{eq:E}
E_{\rm n} = E_{\rm{H}}+E_{\bar{\nu}_e}\,,
\end{equation}
According to the energy–momentum relation and the momentum conservation, $P_{\rm{H}}+P_{\bar{\nu}_e}=0$, Eq.~\ref{eq:E} is then expressed by
\begin{equation}\label{eq:pm}
M_{n} = \sqrt{M_{\rm{H}}^2+P_{\rm{H}}^2}+\sqrt{M_{\bar{\nu}_e}^2+P_{\bar{\nu}_e}^2}\,,
\end{equation}
which yields
\begin{equation}\label{eq:pm1}
\left[M_n -\sqrt{M_{\rm{H}}^2+P_{\rm{H}}^2}\right]^2 = M_{\bar{\nu}_e}^2+P_{\rm H}^2\,,
\end{equation}
and by expanding it, one obtain
\begin{equation}\label{eq:pm2}
\frac{M_{\rm n}^2+M_{\rm H}^2-M_{\bar{\nu}_e}^2}{2M_n}= \sqrt{M_{\rm H}^2+P_{\rm H}^2}
\end{equation}
In the non-relativistic limit for H,
\begin{equation}\label{eq:lim}
\sqrt{M_{\rm H}^2+P_{\rm H}^2}\approx M_{\rm H} + T_{\rm H}\,,\quad {\rm with}~ T_{\rm H} = \frac{P_{\rm H}^2}{2M_{\rm H}}
\end{equation}
Inserting it into Eq.~\ref{eq:pm2}, the expression for the kinetic energy of H, $T_{\rm H}$, then reads
\begin{equation}\label{eq:TH}
T_{\rm H} = \frac{(M_{\rm n}-M_{\rm H})^2-M_{\bar{\nu}_e}^2}{2M_{\rm n}}\,.
\end{equation}
As known, $M_{\rm n}-M_{\rm H}\approx 782.15$ keV, $M_{\rm n}=939.56542$ MeV, and the neutrino energy $M_{\bar{\nu}_e}\ll 1$ eV. Based on these values, the recoil energy of H is about $T_{\rm H}=325.56$ keV. Inversely, from Eq.~\ref{eq:TH}, the mass of a neutrino can be obtained by precisely measuring the kinetic energy of the recoiling nucleus,
\begin{equation}\label{eq:Mnu}
M_{\bar{\nu}_e}\simeq \left[(M_{\rm n}-M_{\rm H})^2 - 2M_{\rm n}T_{\rm H}\right]^{1/2}\,.
\end{equation}
Taking derivative of both sides of Eq.~\ref{eq:TH}, we can find
\begin{equation}\label{eq:dT}
dT_{\rm H} = -\frac{M_{\bar{\nu}_e}}{M_{\rm n}}dM_{\bar{\nu}_e}\,,
\end{equation}
It follows from the prefactor of $M_{\bar{\nu}_e}/M_{\rm n}$ that accurate measurement of $M_{\bar{\nu}_e}$ requires a very high measurement accuracy for $T_{\rm H}$, and, e.g., constraining the neutrino mass down to 0.1 eV requires an accuracy of $T_{\rm H}$ to the level of  $10^{-11}$ eV.

\subsection{Energy Loss of High-Energy Particles in Microcalorimeter Absorbers}\label{B}

Let us first investigate the various possible background signals, which will involve interactions between high-energy particles and experimental materials. The types of particles include mainly high-energy muons, electrons , $\alpha$ particles and $\gamma$ rays, as well as low-energy hydrogen atoms and protons. Neutrinos are also the main secondary particles during the neutron decays, whereas the interactions with the material can be completely neglected due to their extremely high penetration capacity. High-energy muons come from cosmic rays with a typical energy of 4 GeV, and high-energy electrons and lower-energy protons primarily originate from the three-body decay of neutrons, with electron energies $\lesssim782$ keV and proton energy $\lesssim750$ eV. Due to the use of a 1 kV electrode device, proton and electron signals with energies $\lesssim1$ keV are fully screened out and can therefore be safely ignored. Moreover, $\gamma$ rays from neutron sources have a typical energy around MeV, and $\alpha$ particles are produced by the $\alpha$ decay of radioactive nuclei, also with a typical energy of MeV. As mentioned before, the two-body-decay induced hydrogen atoms typically have the kinetic energy of 325.7 eV. For the MeV $\gamma$ rays, the penetration power is known to be extremely high, resulting in an extremely low cross section for direct interactions with the microcalator, so we can also ignore their effects completely. For $\alpha$ particles, due to their relatively weak penetration power, if it comes from the natural radioactive nuclei in the microcalorimeter itself, all their energies will be deposited on the microcalorimeter and thus a detailed analysis of their energy loss in the microcalorimeter absorber is not necessary. Similar to the $\alpha$ particles, the low-energy hydrogen atoms with kinetic energy of 325.7 eV can only penetrate metallic materials of a few nanometers, and all their energies will be deposited in the microcalorimeter, according to the simulation of the~\texttt{SRIM} software~\footnote{http://www.srim.org/}.

Therefore, in the following we will focus on the analysis of the interactions of high-energy muons and electrons with microcalorimeter absorbers. Both muons and electrons are charged particles, and when they interact with materials, the mean rates of energy loss can be given by the Bethe-Bloch formula~\cite{8-Detector-theory}, which reads
\begin{equation}\label{eq:BB}
-\left<\frac{dE}{dx}\right> = \frac{4\pi z^2e^4NZ}{m_0v^2}\left[\ln\left(\frac{2m_0v^2}{I(1-\beta^2)}\right)-\beta^2-\frac{C}{Z}\right]\,,
\end{equation}
which is for a particle with speed $v$, charge $z$ (in multiples of the electron charge $e$) and energy $E$, traveling a distance $x$ into a target with the mean excitation potential $I$ (usually measured through experiments). Here $\beta=v/c$ is the velocity ratio, $N$ is Avogadro constant, $Z$ is the atomic number of the atoms of the absorber material, and $C/Z$ is the shell correction.  $m_0$ is the mass of the incident particle.

According to Eq.~\ref{eq:BB}, the energy loss rate is about 2 MeV per $\rm g/{cm}^2$ for muons with $\beta\rightarrow 1$~\cite{8-Detector-theory}. For 200 nm thick copper and 80 nm thick molybdenum, the energy loss for vertical penetration is about 535 eV. Due to the broad distribution of incident directions of the muons, the energy left in the TES per penetration will be $\ge 535$ eV. Since the lowest signal produced by the muons on the microcalorimeter is about 200 eV higher than the target signal of 325.7 eV, it is easily excluded from signal amplitude and the high-energy muons do not  contribute a continuous background on the target signal.

\begin{figure*}[!htb]
\includegraphics[width=0.8\hsize]{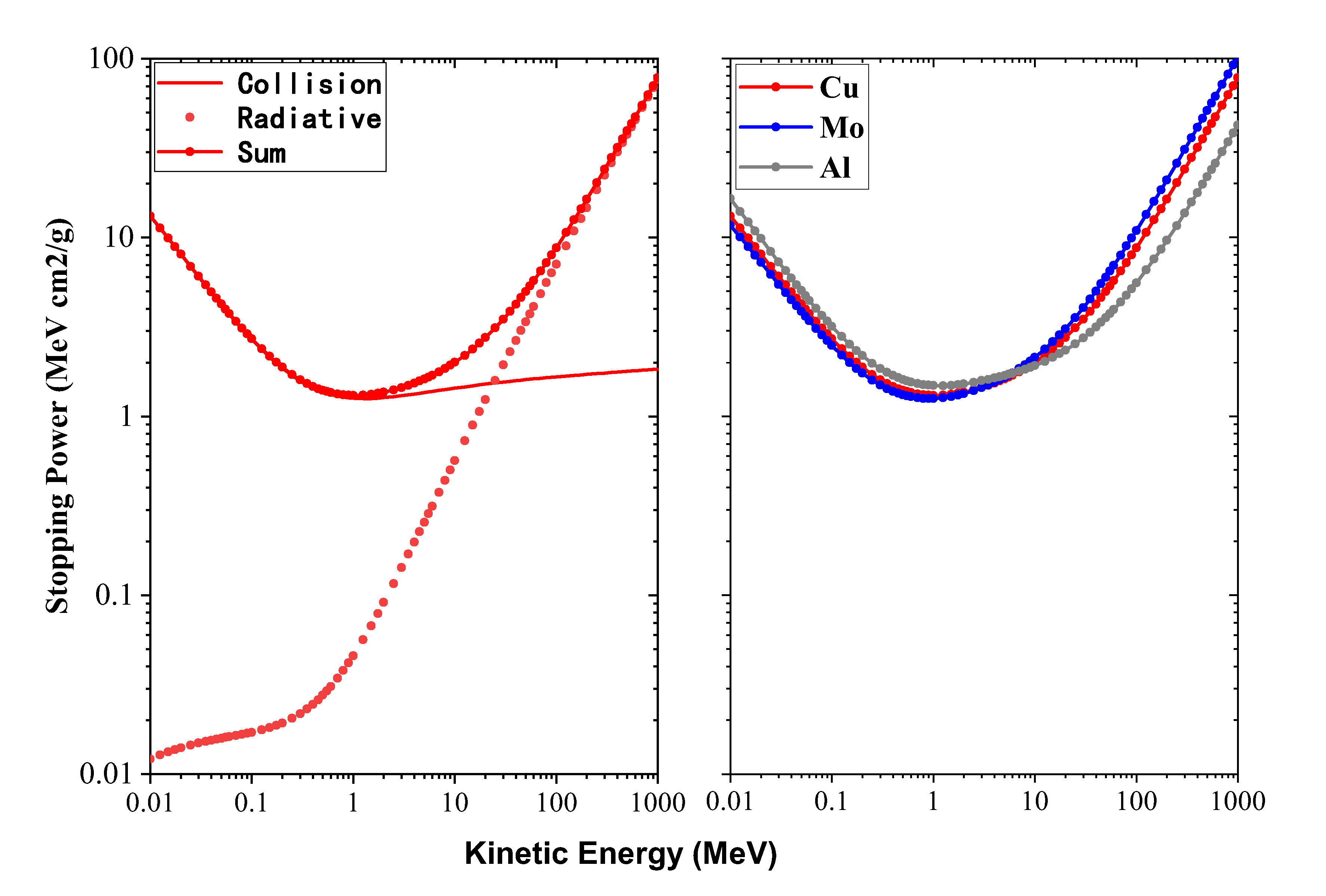}
\caption{{\it Left}: energy loss rate(Stopping Power) of electrons in copper due to collision and radiation at different kinetic energies. {\it Right}: total energy loss rates of electrons in different metallic materials at different kinetic energies.}
\label{fig4}
\end{figure*}

For high-energy relativistic electrons, the energy loss in the interaction with the material varies for its kinetic energy. Since the mass of electrons is much smaller than that of heavy ions, electromagnetic waves are radiated during the acceleration of electrons. Therefore, when the high-energy electrons collide with a material, there is a certain amount of radiated energy loss along with the collision energy loss.

The left panel of Fig.~\ref{fig4} shows the collision and radiated energy loss rates at different electron kinetic energies. As seen, when $E<10$ MeV, the energy loss is dominated by the collisions. The electron kinetic energy from neutron three-body decay is $E <0.78$ MeV, so that we mainly account for the collision energy loss. It can be seen that in the energy range of interest, at the same material thickness, the energy loss of high-energy electrons is rather low, because if the electron energy is high enough to penetrate the material, the high-energy electrons leave only a small fraction of their energies in the detector. In the right panel, the energy loss rates for different metallic materials are shown. As will be mentioned later, molybdenum and copper are the two main elements in our microcalorimeter, and copper and aluminum are the most used elements in the main structure of the refrigerator.

\subsection{Classification and characteristics of neutron sources}\label{C}

Nuclear reactors, radioisotopic neutron emitters and accelerators are the main sources of neutrons, with typical intensities of $10^2$--$10^8$ and $10^8$--$10^{12}$ n$\rm s^{-1}$ for the first two, respectively, and the flux of $10^{10}$--$10^{15}$ n$\rm s^{-1} cm^{-2}$ for the last one.

The neutron emissions are mainly from $\alpha$ neutron, $\gamma$ neutron, and spontaneous fission neutron sources. The nuclear potential barrier increases with increasing the nuclear charge, so the $alpha$ neutron sources generally use light nuclei as targets. Po-Be, Ra-Be and Am-Be sources are the most frequently used neutron $\alpha$ sources. The $\gamma$ neutron source produces neutrons by irradiating $\gamma$ with an energy higher than the binding energy of a target, e.g., Be. The neutron kinetic energy is equal to the photon energy minus the neutron binding energy. Some heavy nuclei can make intense neutron sources, such as californium-252 that emits $2.3\times 10^{12}$ neutrons per gram per second through spontaneous fission. When a few very heavy nuclei absorb a neutron, they become so destabilized that they usually split and eject a few neutrons, which is called nuclear fission. Nuclear reactors thus produce neutron sources through fission nuclei and producing a self-sustaining nuclear chain reaction. Moreover, accelerator-based neutron sources produce neutrons by bombarding specific nuclei with high-energy charged particles produced by accelerators, which has the advantage of high intensity and can obtain monochromatic pulsed neutron beams in a wide energy interval. Neutron sources from accelerators are very intensive, since bombarding the target nucleus with GeV particles will emit multiple neutrons simultaneously. Since accelerator-based source is also pulsed and the target position is well determined, the decay position of the neutrons can be roughly estimated from the neutron velocity. In turn, measurements can be made at specific times and specific locations, allowing for background subtraction for neutron decay measurements.

\begin{figure}[!htb]
\includegraphics[width=0.8\hsize]{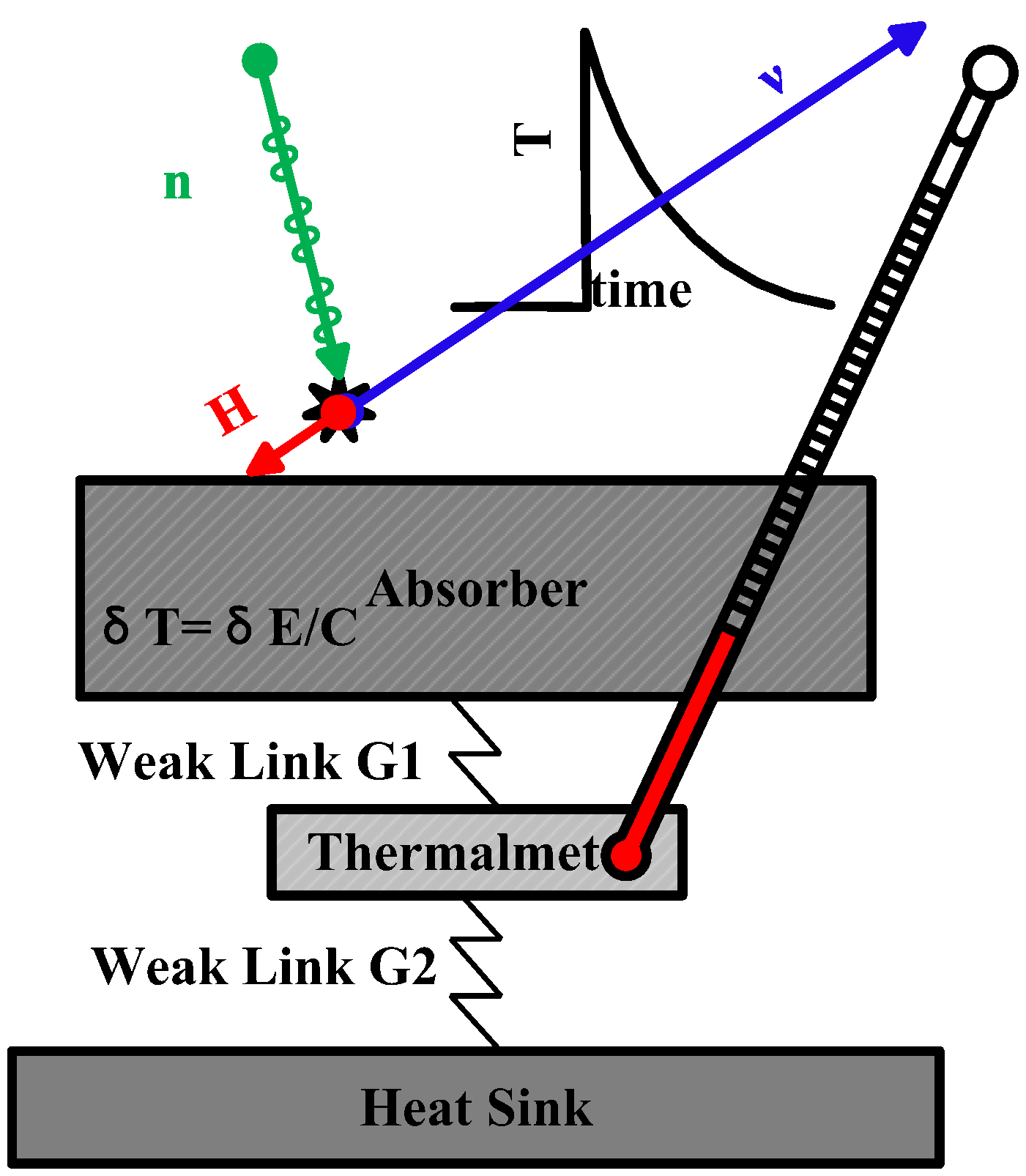}
\caption{Schematic diagram of the structure of microcalorimeter, which directly measures heat signals, insensitive to the type of incident particles.}
\label{fig5}
\end{figure}

\subsection{Working principle and characteristics of microcalorimeter}\label{D}

Microcalorimeter is a new type of detector proposed by D. McCammon et al. in the 1980s~\cite{2-uCal-theory,4-Si-uCal,4-uCal-Rev}, which measures heat changes, with very low energy thresholds and excellent energy resolution.
In principle, any type of particles that is deposited with the same energy and without quenching effects will produce the same signal. In principle, the signal produced by any type of particles remains the same as long as they are deposited with the same energy and without quenching effects. Then, the energy resolution of microcalorimeter, $\delta E_{\rm FWHM}$, is independent of the incident particle energy ~\cite{3-basic-theory} in the linear regime,
\begin{equation}\label{eq:deltaE}
\delta E_{\mathrm{FWHM}} \propto \sqrt{4 k_{\mathrm{B}} T_0^2 C / \alpha_I}\,,
\end{equation}
where $T_0$ is the temperature of absorber, $C$ is the heat capacity of absorber, $\alpha_I$ is the temperature sensitivity of  thermometer.

Moreover, the choice of absorber material for the detector is very flexible, allowing the material to be changed according to the type of particles being detected. Also, since the absorber surface of this detector has the same structure as the inner one, there is no problem of dead layer that occurs in semiconductor detectors, so that it can be used especially for measuring the energies of particles with very weak penetration power.

A microcalorimeter is conceptually a simple device made of four components: an absorber, a thermometer, a weak thermal link and a heat sink, which is shown in Fig.~\ref{fig5}. The absorber is thick enough to stop the particles of interest, but must have the minimum possible specific heat capacity, at cryogenic temperature, to allow for a large temperature rise proportional to the energy deposition of the particles. A weak thermal link to a heat sink is employed to cool a pixel back to its base temperature. The electrical resistance of the thermometer changes inversely as temperature rise, resulting in a voltage signal. As such, the energy of incident particles can be derived from the received electrical signals, along with an appropriate calibration. To balance parameters such as cooling power of the thermostat and energy resolution, a microcalorimeter typically operates at 50 mK or lower. Three classes of microcalorimeters by thermistor technology are frequently used, including semiconductor, transition edge sensor (TES)~\cite{5-TES-uCal}, metallic magnetic calorimeter (MMC)~\cite{6-MMC-uCal}. A TES thermistor has a very high responsivity, with an excellent energy resolution~\cite{4-TES0.75eV}, and a MMC is a dissipation-less system
 that greatly improve some thermal management issues on large detector arrays, enabling excellent repeatability of measurements.

\section{Design Scheme}\label{sec.III}
\subsection{Neutron source requirements}\label{A}
\textbf{Initial kinetic energy of neutrons:}
According to two factors, the width of the energy spectrum and the decay distance of the neutrons from the initial kinetic energy, lower energy neutrons are required for the purpose of this study. Since the neutrino mass in the BoB is completely negligible, almost all of the initial kinetic energy of the neutron is transferred to the hydrogen atom. Since emitted neutrons have an energy broadening, and their energy will be further broadened uniformly upon their interactions with moderator. Therefore, the initial neutrons are not monochromatic, leading to hydrogen atoms with broadening energies. The only way to control the energy broadening of hydrogen atoms is to reduce the initial kinetic energy of neutrons. Since the energy resolution of the detector is intended to be better than 1 eV, it is thus required that the initial kinetic energy of the neutron should be lower than 1 eV. In addition, since the probability of neutron decay near the microcalorimeter is proportional to its side length and inversely proportional to the neutron velocity, the coverage of the microcalorimeter's pixels for neutron trajectories at lower kinetic energies will be better.

For a typical neutron of 4Å wavelength from the Chinese Spallation Neutron Source with a kinetic energy of about 5 meV ($v\simeq 976 {\rm m/s}$), the travel time through a microcalorimeter with the side length of 500 $\mu$m is about 0.512 $\mu$s, and the probability of a single neutron decaying in the path is about $5.8\times10^{-10}$. If an ultracold neutron with an initial kinetic energy of $10^{-7}$ eV were used instead, the probability would increase by about two orders of magnitude.

\textbf{Neutron flux requirements:} with a sufficiently high neutron flux, the single-pixel count rate can be kept around 10 cps (count per second), according to the optimal resolution requirement of the microcalorimeter. According to the predicted two-body-decay branching ratio, the count rate is about $4\times10^{-5}$ cps. In order to accumulate hundreds of counts, one pixel needs to accumulate $2.5\times10^6$ s (about 29 days). This time requirement is relatively high, so when the neutron flux is high enough, it is necessary to consider increasing the single-pixel count rate to obtain more counts.

At present, the number of pixels of microcalorimeters can reach a scale of 100 or more, which means that it takes about 3 days to accumulate a few thousand two-body decay events, so the measurement can be completed in an affordable time, while the flux requirement of ultracold neutrons is about the order of $10^9$--$10^{10}/\rm cm^2$. In addition, neutron sources are often accompanied by $\gamma$ rays, which will be briefly discussed later.

\begin{figure*}[!htb]
\includegraphics[width=0.8\hsize]{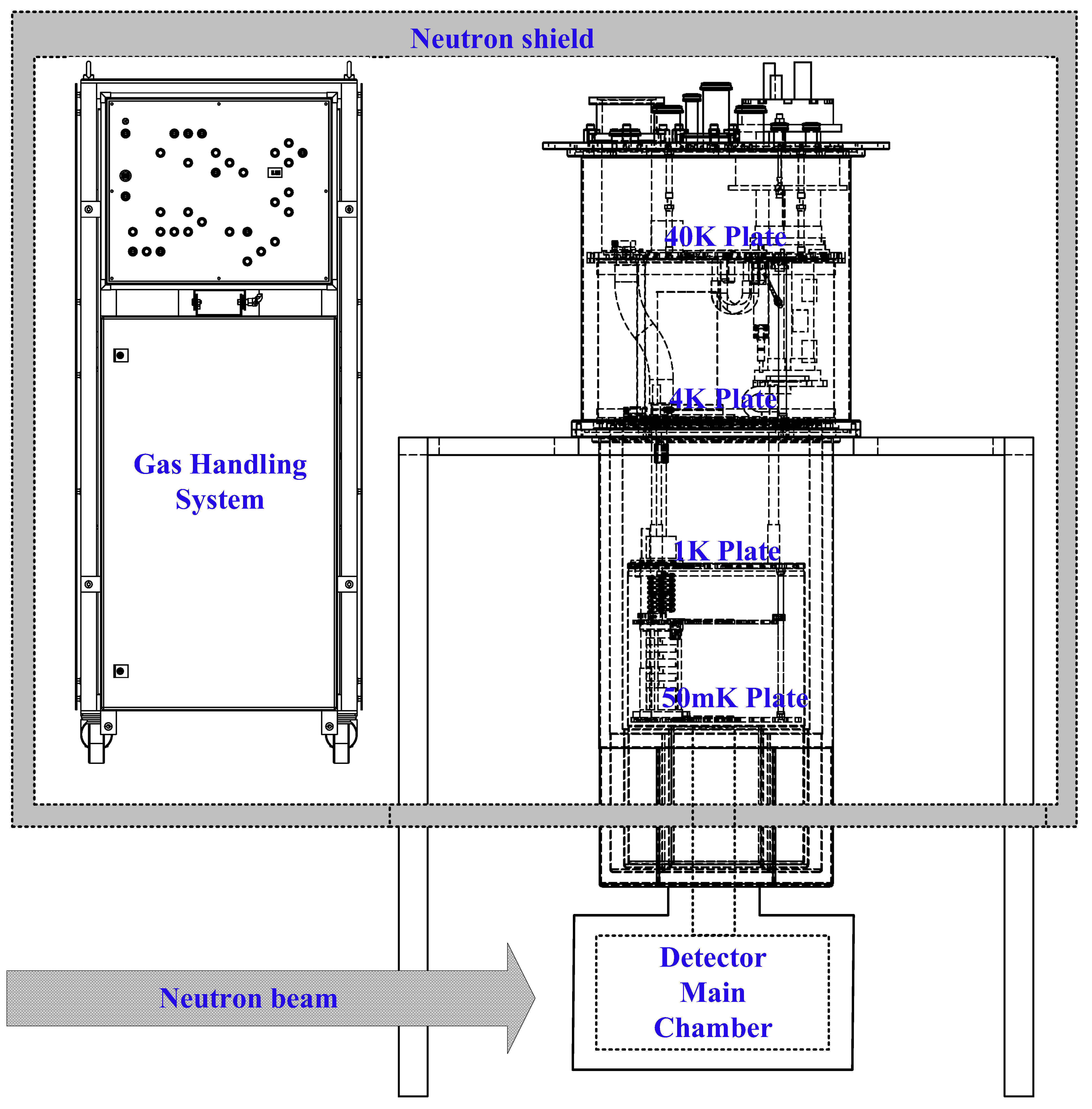}
\caption{Schematic diagram of the $^3$He/$^4$He dilution refrigerator. The refrigerator provides continuous cooling to sufficient low temperatures for the BoB measurement, while a shielding layer of neutrons to protect the $^3$He is necessarily required.}
\label{fig6}
\end{figure*}

\subsection{$^3$He related issues in refrigerator}\label{B}

The microcalorimeter needs to reach a temperature of 50 mK or even lower to obtain a sufficiently high energy resolution. At present, adiabatic demagnetization refrigeration (ADR) or dilution refrigeration (DR) is commonly used to obtain such low temperatures.

The ADR uses the magnetocaloric effect caused by changing the external magnetic fields to change the state of the magnetic material, with the minimum operating temperature of 40 mK and without the influence of gravity. The ADR has high refrigeration efficiency, allowing for miniaturization and weight reduction. However, the ADR cannot continue cooling and prevent temperature rise during the magnetization process as it have to operate with a repeated magnetization and demagnetization process. Since the ADR must be terminated when the temperature rises to about 100 mK in the magnetization process, it is not available for experiments that require long periods of time, such as the BoB measurements.

The DRs utilize the properties of $^3$He and $^4$He mixtures to achieve cooling, which have the advantages of lowest temperature down to several mK, continuous cooling, and high cooling power. The operation of a DR depends on gravity to keep $^3$He and $^4$He in their correct chambers and the size is typically quit big but manageable. In particular, it can operate continuously for a long time in the range of 10--300 mK, which is suitable for our needs in this study. However, $^3$He is a strong absorber of neutrons, reacting by absorbing neutrons and producing a proton and a triton. Such reaction will lead to the failure of DR. Therefore, in order to shield the neutrons to avoid this problem, the $^3$He loop needs to be covered with a protective layer of resin for a sustainable operation of DR.

\begin{figure*}[!htb]
\includegraphics[width=0.8\hsize]{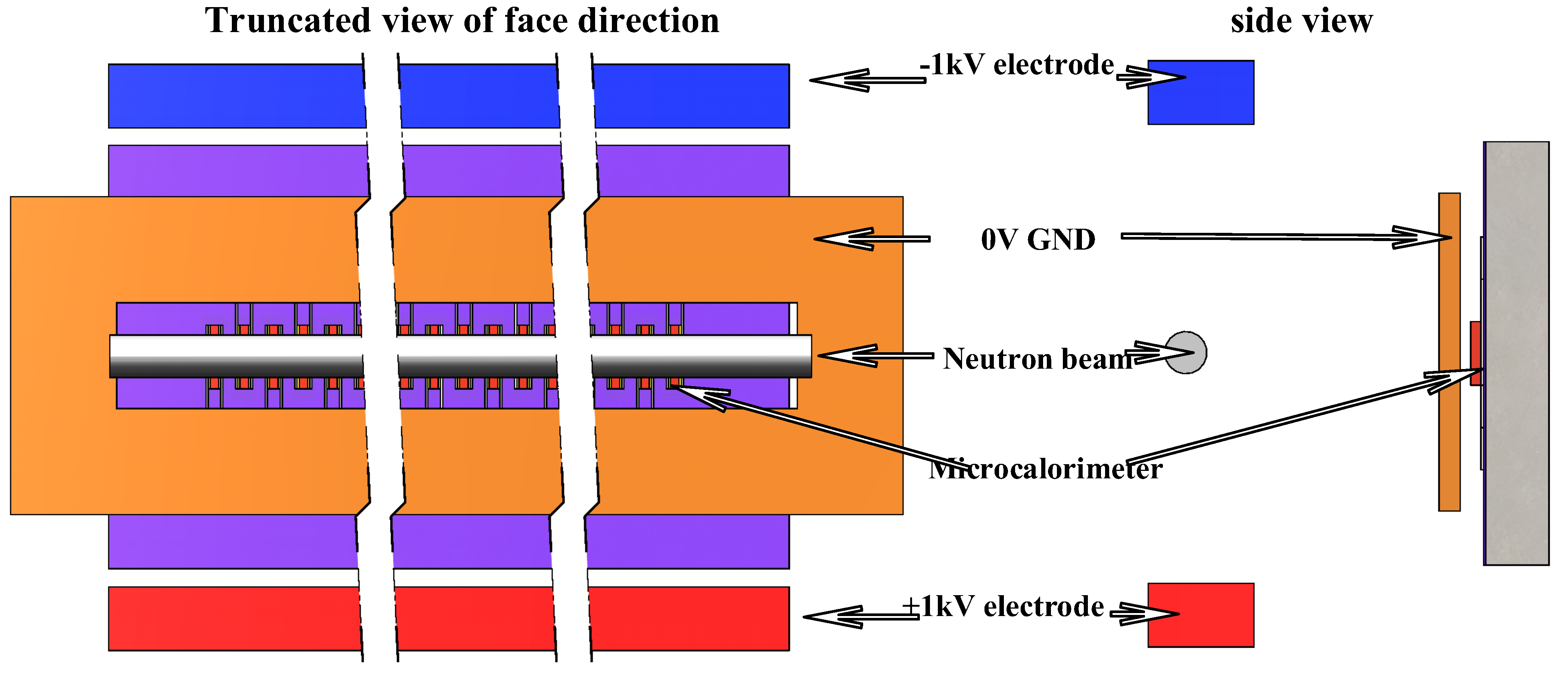}
\caption{Electrode layout scheme for voltages of $\pm 1$ and 0 kV.}
\label{fig7}
\end{figure*}

\subsection{Eliminating impacts of protons}\label{C}
The three-body-decay produced protons will contribute a strong continuum background of about $E<750$ eV against the signal of hydrogen atoms from the BoB process. Therefore, such background have to be subtracted completely.
It is suggested to add the high-voltage of $\pm1$ kV at the upper and lower ends parallel to the neutron beam, respectively, to remove all charged particles with $E<1$ keV. Also, the front of the microcalorimeter needs to be equipped with a shield to screen the electric field by the high voltages.

\subsection{Impacts of high-energy particles}\label{D}

In measurements, the radiation background of the environment has some impacts on the determination of the hydrogen signal, which will be briefly discussed as follows. In the energy spectrum, especially near the target energy of 325.7 eV, there are mainly two dominated components: the continuum energy spectrum and the elemental spectral lines.

The sources of the continuum spectrum around 325.7 eV originate mainly from the following scenarios, including 1) energetic charged particles penetrating the microcalorimeter, 2) scintillation lights shining on the microcalorimeter, where the lights are produced from the interactions of high-energy particles with the material surrounding the microcalorimeter, and 3) secondary particles with energy levels of a few hundred eV, from energetic particles interacting  with the material around the microcalorimeter.

Furthermore, the characteristic spectral lines of elements near 325.7 eV are in principle negligible. This is because the elements that make up the microcalorimeter itself and its surrounding mainly include copper, molybdenum, carbon, aluminum, gold, and silicon, which have no characteristic lines in the range of 10--325.7 eV. Therefore, only the effects of the continuum energy spectrum needs to be considered. Cosmic rays and naturally occurring radioactive radiations are the main sources of the radiation background, and in addition, if spallation neutron sources are used to provide neutrons, the neutron beam is often accompanied by MeV $\gamma$ rays.

\textbf{Impacts of cosmic rays:}
at ground level, muons are the most abundant particles produced from the interactions of primary cosmic rays at the top of the atmosphere, with energies of the order of GeV and the typical flux of about 1 $\rm s^{-1}cm^{-2}$. When mouns passing through the absorber, due to its $\mu$m scale and the extremely low ionization cross section of muons, the energy residuals in the absorber are expected to be the order of $<1$ ev. The residuals is thus so different from the target energy that their effects are fully negligible. However, the muons will hit the interior of the thermostat to produce scintillation lights, and the microcalorimeter will be able to respond to such lights. Therefore, the surrounding components of TES should be minimized in designing the structure. Also, these lights can be effectively reduced by light absorbing coating.

\textbf{Impacts of naturally-occurring radionuclides:}
the absorber contains about $10^{17}$ atoms, and if the main component is copper, assuming 1ppb of natural radioactive contaminants, there are about $10^{8}$ radionuclides. According to their typical half-life of about $10^8$  years, there is about one decay event per year, thus with no contamination on the BoB measurement.

\textbf{Impacts of  $\gamma$ rays:} with a very low possibility of interactions with the absorber ($\mu$m thickness), $\gamma$ rays have an extremely high penetration power and they do not interact directly with microcalorimeters. Similar to the cosmic muons, the effects of the $\gamma$-ray induced scintillation lights are also expected to be negligibly small.

\subsection{Heat release from absorption of hydrogen atoms}\label{E}
When the hydrogen atoms interact with the absorber, some chemical energies are released at the eV level, leading to a final microcalorimeter-measured heat signal slightly higher than 325.7 eV. The shift of the spectral lines at the eV level does not affect the confirmation of the BoB mode, although the spectral line shifts of the order of eV have a significant impact on inferring the physical properties of neutrinos. In principle, this shift would be well corrected by theoretical calculations and experimental calibrations, and we will leave this for future investigations

\subsection{Design of the Microcalorimeter and Test System}\label{F}
\subsubsection{Parameters requirements of the Microcalorimeter}
\textbf{Energy resolution:}
In Sect.~\ref{A}, the use of ultracold neutrons with kinetic energies below    $10^{-7}$ eV, would hardly lead to broadening of the energy spectrum of hydrogen atoms. When hydrogen atoms interact with materials, a certain amount of chemical energy will be released, and the hydrogen atoms will eventually bind to the absorber in the form of covalent or ionic bonds. The binding process will certainly yield energy fluctuations, and thus lead to an intrinsic broadening of the amplitude of the total heat signal, with width of $\ll 1$ eV . In this sense, the energy resolution of the microcalorimeter needs to be of the order of eV or even sub-eV. Meanwhile, the higher the energy resolution of the detector, the more the number of background signals can be subtracted from the continuum energy background radiation, thus also requiring the energy resolution to be achieved as high as possible. From Eq.\ref{eq:deltaE}, the high energy resolution corresponds to a small heat capacity $C$, thus demanding a smaller microcalorimeter pixel size.  However, this conflicts with the requirement of a large absorption surface for the BoB measurement. We will perform a more detailed investigation on the effective area in a moment.

\textbf{Count rate:}
provided that the neutron flux of the neutron source is sufficiently high, the higher the single-pixel count rate of the microcalometer will of course lead to the higher the statistics achievement, thus improving the measurement accuracy and reducing measurement time. For conventional microcalometer covering energy below 10 keV, typical pulse recovery times are a few ms. As neutron decay is a quasi-random signal in time domain, there is no significant signal accumulation for single-pixel count rates of 50-100 cps. Since the target energy is only 325.7 eV, it is more appropriate to choose a smaller heat capacity $C$ of the microcalorimeter. According to the time estimate $\tau=C/G$, where $G$ is the thermal conductivity and generally a relatively constant, the typical pulse recovery time of the microcalorimeter can be on the order of sub-milliseconds, i.e., in principle the design allowing for single-pixel count rates in the range of 500-1000 cps. However, limited by the ultracold neutron flux from the current spallation neutron source, a count rate of 50 cps is sufficient for the measurement.

\textbf{Effective area:}
As mentioned above, the low flux of the current ultracold neutron source requires increasing the effective area to improve the collection efficiency of hydrogen atoms. The heat capacity $C$ of the absorber is proportional to its area for a fixed absorber thickness. Eq.~\ref{eq:deltaE} indicates that $C$ is inversely correlated with the energy resolution. As such, a large effective area help for achieving high energy resolution. Currently, the effective area of microcalorimeters with energy resolution below 1 eV is about $200\times200 \mu{\rm m}^2$~\cite{4-TES0.75eV}. To compromise the effective area and energy resolution, a larger effective area can be obtained by increasing the number of pixels, but this demands to multiplex the readout system .

\subsubsection{Parameter requirements for the microcalorimeter}

As mentioned in Sect.~\ref{D}, the TES microcalorimeter can achieve the highest energy resolution~\cite{4-TES0.75eV}, which is the most important parameter for the BoB measurement. Compared with the MMC which has a better linearity, the TES can achieve a higher multiplexing factor for SQUID, which gives a great advantage in expanding the total effective area. In short, TES microcalorimeter would be the best choice for measuring neutron two-body decay. At present, the achievement of the highest energy resolution is based on a molybdenum-copper bilayer, and therefore we will focus our investigation on it in the following.

\begin{figure}[!htb]
\includegraphics[width=1\hsize]{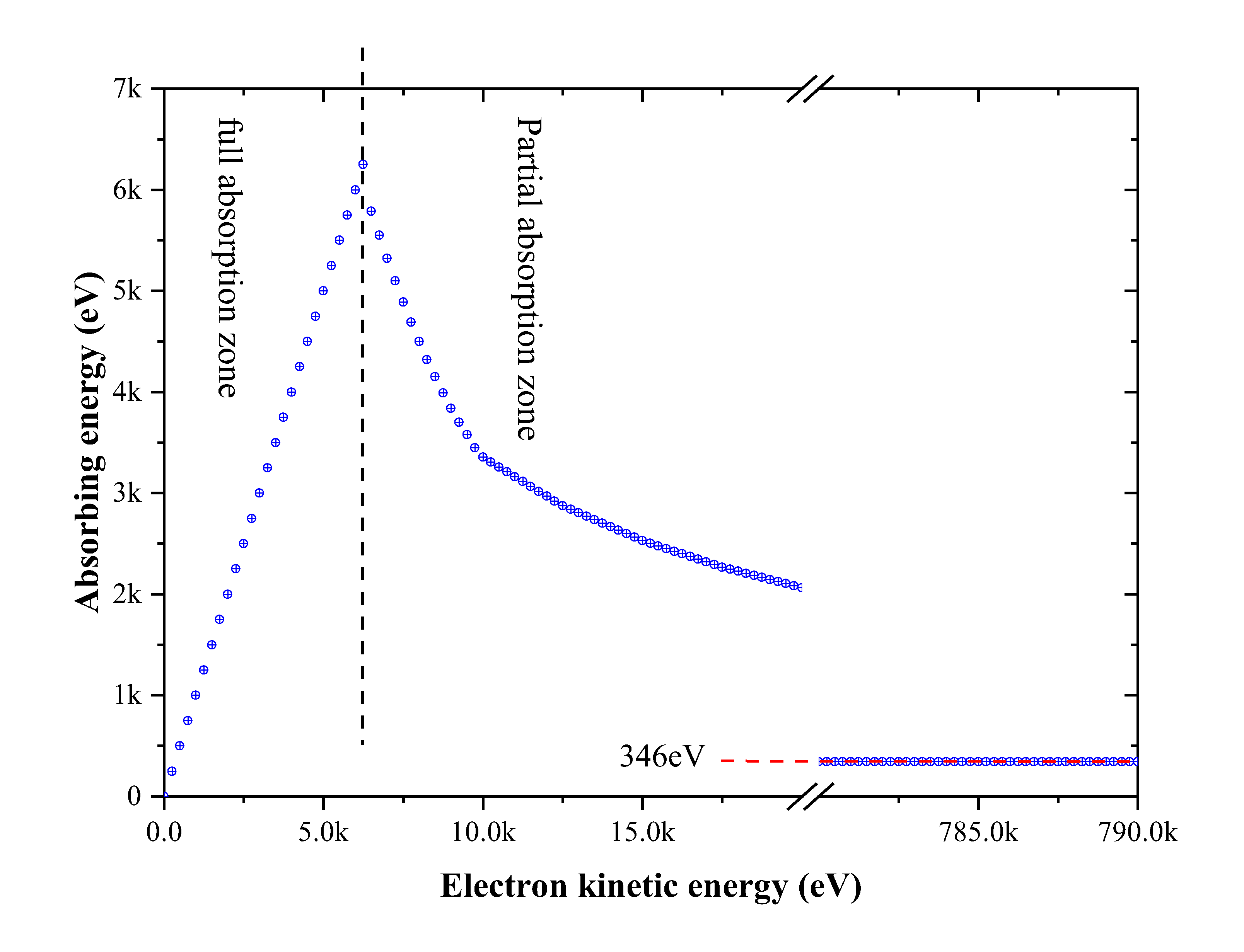}
\caption{Dependence of the absorbed energy of two-layer films consisting of 200 nm thick copper and 80 nm thick molybdenum on the incident electron energy.}
\label{fig8}
\end{figure}

\subsubsection{Structure design of the Microcalorimeter}
\textbf{Absorber Design:}
At present, TES for particle detection mostly uses molybdenum-copper, molybdenum-gold and titanium bilayer films, or an aluminum-manganese alloy film, etc. Due to the poor penetration power of the hydrogen atoms, the thermometer itself can capture it quite well, the TES-microcalorimeter for BoB does not require the use of an additional absorber. According to the calculation, the initial kinetic energy of hydrogen atoms is so small that a material with a thickness of a few nanometers can dissipate all its energy.
Taking the molybdenum-copper bilayer TES as an example, the transition temperature below 100 mK would be obtained with a thickness of about 80 nm for molybdenum and about 200 nm for copper, which itself, in other words, can meet the measurement requirements as an absorber.

In the reality, the protons from three-body neutron decay  will have been eliminated by the 1 kV electrode, but the electrons with energy higher than 1  keV will hit the microcalorimeter and produce a signal. For electrons with energies below 6 keV, their kinetic energy is not sufficient to penetrate the molybdenum-copper bilayer, so electrons with energies below this are all deposited into the microcalorimeter. For electrons with energies higher than 6 keV, their kinetic energies are so high that they will penetrate the microcalorimeter, leaving some energy being absorbed. According to the energy loss rates of high-energy electrons (see the right panel of Fig.\ref{fig4} based on Eq.~\ref{eq:BB}), the resulting energy absorption below 782 keV in the molybdenum-copper bilayer film are derived and shown in Fig.~\ref{fig8}.
The residual energy of electrons of 782 keV is about 346 eV, about 20 eV shift from the target spectral line of 325.7 eV, thus without biasing our BoB measurement.

\begin{figure*}[!htb]
\includegraphics[width=1\hsize]{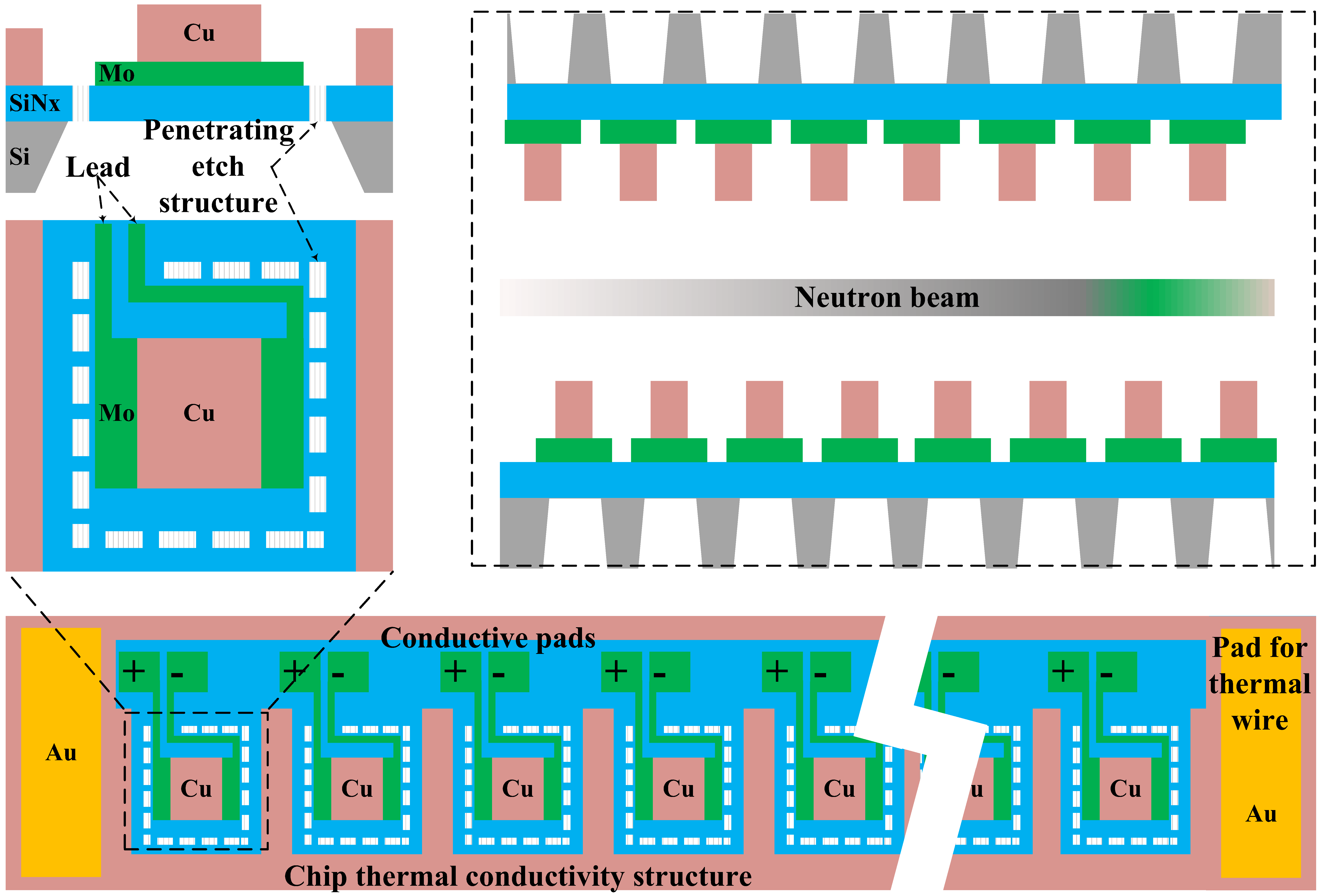}
\caption{Schematic diagram of {\bf XXXX}. {\it Upper right}: the microcalorimeter chip is placed parallel to the neutron source beam and each pixel is distributed in one dimension on the chip. According to the design, each pixel contains molybdenum wires, molybdenum-copper TES, and copper for heat conduction.}
\label{fig9}
\end{figure*}

\textbf{Structure design of the microcalorimeter:}
To efficiently utilize the neutron source beam, the microcalorimeter chip needs to be placed parallel to the beam direction, and each pixel needs to be arranged in one dimension on the chip. In doing so, it helps to improve the process uniformity between pixels. For individual pixels, molybdenum-copper bilayer films are used as TES with a thickness of about 80 nm for molybdenum and about 200 nm for copper. The transition temperature of single-layer molybdenum is about 900 mK, well above the operating temperature, so molybdenum can be used as a wire, reducing micro-machining steps and thus improving process consistency. A single layer of copper is used as a thermal conductor to ensure that the heat induced by the high-energy particles is conducted out timely. In order to weld the heat sink of the refrigerator with the heat conduction wire, a solder pad is arranged at each end of the chip.

\subsection{Multiplexing method of readout electronics for TES-type microcalorimeters}\label{G}
Increasing the number of pixels in a microcalorimeter can directly increase the total effective area and significantly reduce the measurement time. The readout electronics currently in TES microcalorimeters commonly use the  superconducting quantum interference device (SQUID). However, the SQUID is costly, so a SQUID needs to read out multiple TES to increase the number of pixels. Four SQUID multiplexing techniques currently read out large TES arrays in common use: time-division multiplexing (TDM), frequency-division multiplexing (FDM), a code-division multiplexing (CDM) and microwave SQUID multiplexing (uMUX). The first three all use a DC-SQUID amplifier, converting a current change caused by the incident signal into a magnetic flux change and coupling it into a SQUID current amplifier to amplify the output. TDM is the most frequent and established multiplexed readout technology for TES array detectors at present. The TDM is fully compatible with a single-pixel TES, and it is easy to switch between it and TES array through a back-end control circuit. uMUX couples the SQUID current to SQUID loaded microwave resonators, with each resonator tuned to a different frequency, allowing the low power TES current to modulate a higher power RF signal. The uMUX technology will potentially enable ultra large-scale TES array detector readout circuits with 100,000 pixel level, playing a key role in the next-generation multiplexing readout.

\section{Summary and Outlook}\label{sec.IV}
In this paper, a microcalorimeter-based scheme for measuring the two-body neutron decay with extremely high energy resolution is proposed. The signal of the two-body neutron decay is expected to be two single-energy spectral lines at 325.7 eV and 335.9 eV, while it will be accompanied by continuum energy spectra of electrons and protons arising from the neutron three-body decay. From the estimation, this background contamination can be significantly suppressed by adding high-voltage electrodes of $\pm1$ kV. Also, muons and electrons and other particles have potential impacts on the target spectral line measurements. However, we find that the signals produced by these particles are significantly different from the target spectral lines and thus can be easily excluded. Then, the requirements of the BoB measurement for the neutron source are discussed. This experiment would ideally use an ultracold neutron source. It also requires a neutron flux of the order of $10^9$--$10^{10}$ so that the count rate collected by the absorber can reach about 10 cps, equivalent to the accumulation of thousands of neutron two-body decay events within three days. Finally, the energy resolution, count rate and other information required for the BoB are provided. Based on the microcalorimeter with the TES of molybdenum-copper bilayer film, the choice of absorber material and structural design are investigated. A readout scheme for electronic multiplexing of TES is also given, which can be used to improve the measurement sensitivity and reduce the cost.

1). {\it The primary objective} of this measurement scheme is to efficiently determine the existence of the bound state of neutron decay using the energy spectrum-based microcalorimeter, thus promising to confirm for the first time the theoretical hypothesis never observed so far.

2). {\it The next objective} is to measure the branching ratio of the two-body and the three-body decays as well as the population ratios of the ground state (1s) and the excited ones ($n$s), and check whether these ratios are consistent with the theoretical expectations. The difficulty of completing such measurements does not increase significantly.


3). {\it An ambitious and long-term objective} is the direct measurement of the neutrino mass, $M_{\bar{\nu}_e}$. If the kinetic energy of $H$ can eventually be measured with accuracy to the order of $10^{-11}$ eV, the mass of the electron neutrino can be obtained. To obtain a high confidence level, this requires a tremendous amount of measurements. Therefore, it demands a massive expansion of the number of pixels in the microcalorimeter, but which is very costly.

In addition to ground-based particle physics experiments, cosmological observations currently provide important complementary information on neutrino masses, placing the strongest upper limit on neutrino mass. The large number of Big Bang neutrinos affects the evolution of the universe and provides an excellent playground for studying their physical properties over the cosmic scale of tens of billions of light years. Theoretically, the observable effects of cosmology depend mainly on the total mass of neutrinos. In the standard $\Lambda$CDM model, massive neutrinos change the expansion rate of the cosmic background, affecting the anisotropy of the cosmic microwave background (CMB) and its angular power spectrum. Meanwhile, as the universe expands, mass neutrinos, after becoming non-relativistic particles, still have large thermal velocities and strong diffusivity, which will smooth out the small-scale perturbations in the matter density, and cause a suppression on the small-scale matter power spectrum.

Thus, the total neutrino mass affects the large-scale structure of the universe and changes the gravitational lensing effect of the CMB. Currently, combining the Planck CMB data and the large-scale galaxy surveys, astronomical observations give an upper limit of 0.12 eV (95\% C.L.) for the total neutrino mass~\cite{Planck:2018vyg}. The next generation of sky observations (e.g., CMB-S4~\cite{CMB-S4:2016ple}, DESI~\cite{Font-Ribera:2013rwa}) will further improve the neutrino mass constraint, and the measurement error is expected to be narrowed to 0.03 eV (95\% C.L.). Thus, in the next decade, the combination of ground-based particle physics experiments and astronomical observations is expected to accurately determine the absolute masses of neutrinos and answer major physics questions such as mass ordering.

\end{document}